\documentclass[3p,times]{elsarticle}
\usepackage{url,epsfig,textcomp,epstopdf,color}
\usepackage{amsbsy,amsmath, amssymb,comment,dsfont,amsthm}
\usepackage{mathtools,algorithm,algorithmic,multicol,enumitem}
\usepackage{lineno}
\modulolinenumbers[5]
\newtheorem{definition}{Definition}\newtheorem{theorem}{Theorem} \newtheorem{corollary}{Corollary} \newtheorem{lemma}{Lemma} 
\newcommand{\bth}{\begin{theorem}}\newcommand{\ethe}{\end{theorem}} \newcommand{\bpr}{\begin{proof}}\newcommand{\epr}{\end{proof}} \newcommand{\ble}{\begin{lemma}}\newcommand{\ele}{\end{lemma}} \newcommand{\bco}{\begin{corollary}}\newcommand{\eco}{\end{corollary}}
\newcommand{\bde}{\begin{definition}}\newcommand{\ede}{\end{definition}}
 \newcommand{\ci}{~\cite}

\usepackage{caption, subcaption} 
\journal{Journal of Theoretical Computer Science}

\begin{document}
		
\begin{frontmatter}
\title{Oblivious Sorting and Queues}	
\author{Johannes Schneider}
\address{University of Liechtenstein, Vaduz, Liechtenstein}
\ead{johannes.schneider@uni.li}

\begin{abstract}
\noindent We present a deterministic oblivious LIFO (Stack), FIFO, double-ended and double-ended priority queue as well as an oblivious mergesort and quicksort algorithm. 
Our techniques and ideas include concatenating queues end-to-end, size balancing of multiple arrays, several multi-level partitionings of an array. Our queues are the first to enable executions of pop and push operations without any change of the data structure (controlled by a parameter). This enables interesting applications in computing on encrypted data such as hiding confidential expressions. 
Mergesort becomes practical using our LIFO queue, ie. it improves prior work (STOC '14) by a factor of (more than) 1000 in terms of comparisons for all practically relevant queue sizes. We are the first to present double-ended (priority) and LIFO queues as well as oblivious quicksort which is asymptotically optimal. Aside from theortical analysis, we also provide an empirical evaluation of all queues.
\end{abstract}
\end{frontmatter}

\linenumbers

\section{Introduction}\label{sec:intro}
Advances in computing on encrypted data such as Fully Homomorphic Encryption (FHE) and secure multi-party computation (SMC) might make outsourcing computation securely practically feasible. Memory access must also be secured. For example, accessing the $i$-th element of an array of length $n$ needs $O(1)$ operations on RAM machines. But for a program running on encrypted data, the same access mechanism reveals access patterns. The knowledge of seemingly simple access patterns can help to disclose sensitive information such as stock trading patterns\ci{pin10} or encryption keys\ci{isl12}. A simple solution requires to access all array elements requiring $O(n)$ instead of $O(1)$ time. Oblivious RAM (ORAM) secures memory access more efficiently using multiple parties. Often relying on more than one party is not desirable. Current solutions for oblivious data structures also do not hide (high level) operations, which makes them unsuitable for omnipresent `if-then-else' statements with private conditions and queue access in branches. Evaluating a confidential expression, keeping data as well as the expression itself secret, becomes straight forward using our LIFO queue and known techniques for computing on encrypted data. Such a scenario is important for cloud computing, ie. a cloud provider might host data for customers, which run their own analytics functionality. The customers wish to keep their data and algorithms private -- in case of industrial automation an algorithm often means a mathematical expression on time-series sensor data.\footnote{In fact, a request from industry motivated this feature.} To summarize, the main contributions are:
\begin{enumerate}[topsep=0ex]
	\setlength\itemsep{-0.2em}
	\item We present oblivious LIFO, FIFO and double-ended (priority) queues. The amortized overhead of an operation on the LIFO queue is O($\log n$) in the maximal length $n$ of the queue. Prior LIFO queues (based on priority queues\ci{tof11}) required $O(\log^2 n)$.	
	For a wide range of applications such as the producer-consumer problem in a streaming context our FIFO queue has only O($\log n$) overhead which improves prior work\ci{tof11} by a factor $\log n$. We are the first to introduce double-ended queues. Our double-ended queue needs O($\log^2 n$).
	\item We are the first to derive oblivious data structures to support push and pop operations that might not alter the stored elements (depending on a parameter).
	\item Our deterministic mergesort algorithm improves on\ci{goo14} for all relevant list sizes, eg. by two orders of magnitude for sorting of 10 billion elements. 
	\item We state the first oblivious quicksort algorithm. It is asymptotically optimal. The Monte Carlo algorithm succeeds with high probability, ie. $1-1/n^c$ for an arbitrary constant $c$. 
\end{enumerate}

\subsection{Overview of Technique}
We structure the array representing the queue in subarrays (SA) of increasing size. A SA might be itself a queue. SAs are organized into parts that are merged and split if they are shifted between different SAs. Moving of elements between SAs can cause some of the push and pop operations to require linear run-time in the maximal queue length. But the time is amortized across many operations so that the average overhead is only (poly)logarithmic. Moving of parts between SAs happens based on the number of pops and pushes. It is not dependent on the data held in the queue. We develop a deterministic calling pattern that does not require knowing the number of stored elements in a queue. This allows to hide the number of operations together with another idea: We permit the pop and push of a special (empty) element that does not alter the number of stored elements in the data structure. Put differently, this disguises whether an operation on the data structure changed the stored elements or not. Furthermore, to ensure efficient access to both ends of a queue, eg. as needed for FIFO and double ended queues, we concatenate two ends of a (LIFO) queue. 

\subsection{Outline}
We first discuss our model and some notation (Section \ref{sec:pre}).   The main data structures are given in Section \ref{sec:lifo} (Stack) with a detailed explanation of core ideas and analysis, Section \ref{sec:fifo} (FIFO) and Section \ref{sec:array} (double-ended queue). Detailed case studies are given in Section \ref{sec:app} after explaining the technique thoroughly. This includes an explanation how obliviousness (and operation hiding) helps in securing code. Performance evaluation can be found in Section \ref{sec:eval}. 

\section{Preliminaries and Limitations} \label{sec:pre}
We assume knowledge of an upper bound on the maximal number of elements $n$ that can be kept in the data structure, ie. a queue is represented by an array of fixed size $n$. This assumption is common for oblivious data structures. Adjusting the size of the data structure exactly to the actual number of elements is impossible since our goal is to conceal the number of elements contained in the queue. Our queues support two operations: Push (allowing empty elements) and Pop (allowing conditional popping).
For obliviousness we proved an analogous definition as \ci{gol87Obl}. Essentially, obliviousness implies that memory access patterns are the same for any input.
\bde \label{def1}
A data structure is \emph{oblivious} if the sequence of memory access only depends on the number of push and pop operations. A sorting algorithm is oblivious if the sequence of memory accesses is the same regardless of the input. 
\ede

We use a special (empty) element ``\texttt{$\varnothing$}'' also denoted by a dash `$-$` indicating that an element in the queue is unused.  Its bit representation must be different from any data item stored in the queue.
We use variants of compare and exchange operations. The simplest form takes as input a binary bit $b$ and two variables $A$ and $B$. It assigns $A:=B$ if the bit $b$ is 1, otherwise $A$ is not changed, ie. it computes $A := b\cdot B + (1-b)\cdot A$. The compare-exchange-and-erase $CmpExEr(b,A,B)$ performs a compare and exchange as described and, additionally, it might erase $B$, ie. it sets variable $B$ to $\varnothing$ if $b$ is 1 and leaves it unchanged otherwise (see PseudoCode $CmpExEr$ in Algorithm \ref{alg:lifo}).
For the analysis we distinguish between input sensitive operations involving parameters of the push and pop elements as well as data of the queue and operations that do not directly depend on any input data (but potentially on the number of operations). The motivation is that for secure computation these distinctions are meaningful, since the former correspond to (slower) operations on encrypted data. For our algorithms input sensitive operations always dominate the time complexity -- even when using non-encrypted data. They are split into elementary operations (+,-,$\cdot$), called \emph{E-Ops}, and comparisons \emph{C-Ops}, which are composed of elementary operation. The distinction is motivated since comparisons are used to measure performance of sorting algorithms. For encrypted operations, comparisons might have different time complexities, eg. for SMC such as \cite{sch15a} it is not clear how to perform a comparison in less than $\Omega(n_b\cdot E-Ops)$ time, where $n_b$ is the number of bits of a compared number.

\section{LIFO (Stack)} \label{sec:lifo}
For a Last-In-First-Out (LIFO) queue (also called Stack) a pop operation returns the most recently pushed element onto the data structure. 
To ensure obliviousness we access the same array elements independent upon the data contained in the queue. Our queue always accesses the first element. A newly pushed element is stored in the first position of the array. This implies that upon every insertion, we must shift elements to the right to avoid overwriting of a previously inserted element. It is easy to shift the entire queue to the right but this requires linear run-time. To improve efficiency, we logically split the array representing the queue into subarrays (SAs) of exponentially growing size. We only shift parts of size at most $2^k$ of a SA after every $2^{k}$ push or pop operations.

\begin{figure*}[htp]
	\centering	\centerline{\includegraphics[width=\linewidth]{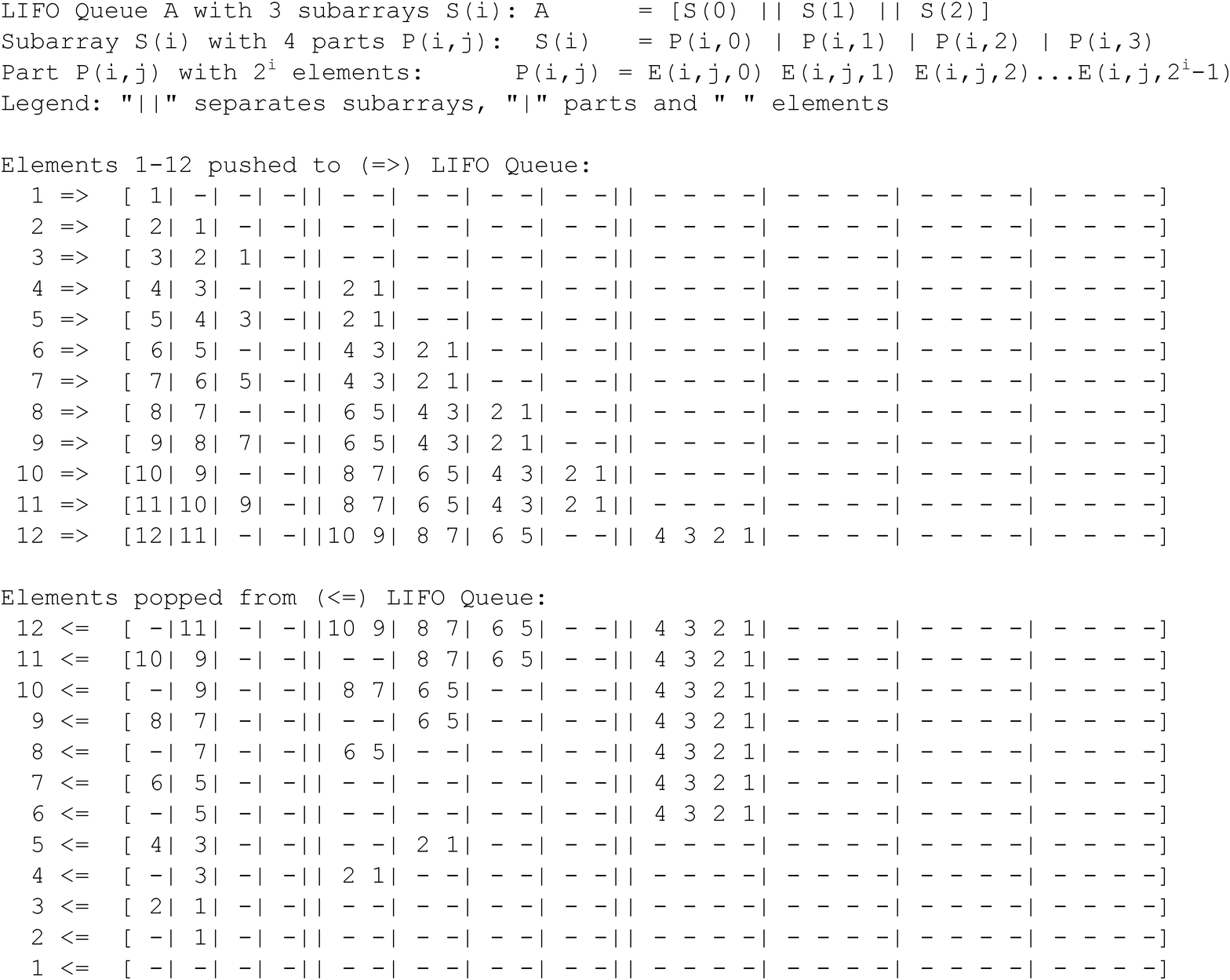}}
	\caption{Pushes and pops onto a LIFO queue}
	\label{fig:basic}
\end{figure*}

\begin{figure*}[htp]
	\centering	\centerline{\includegraphics[width=0.8\linewidth]{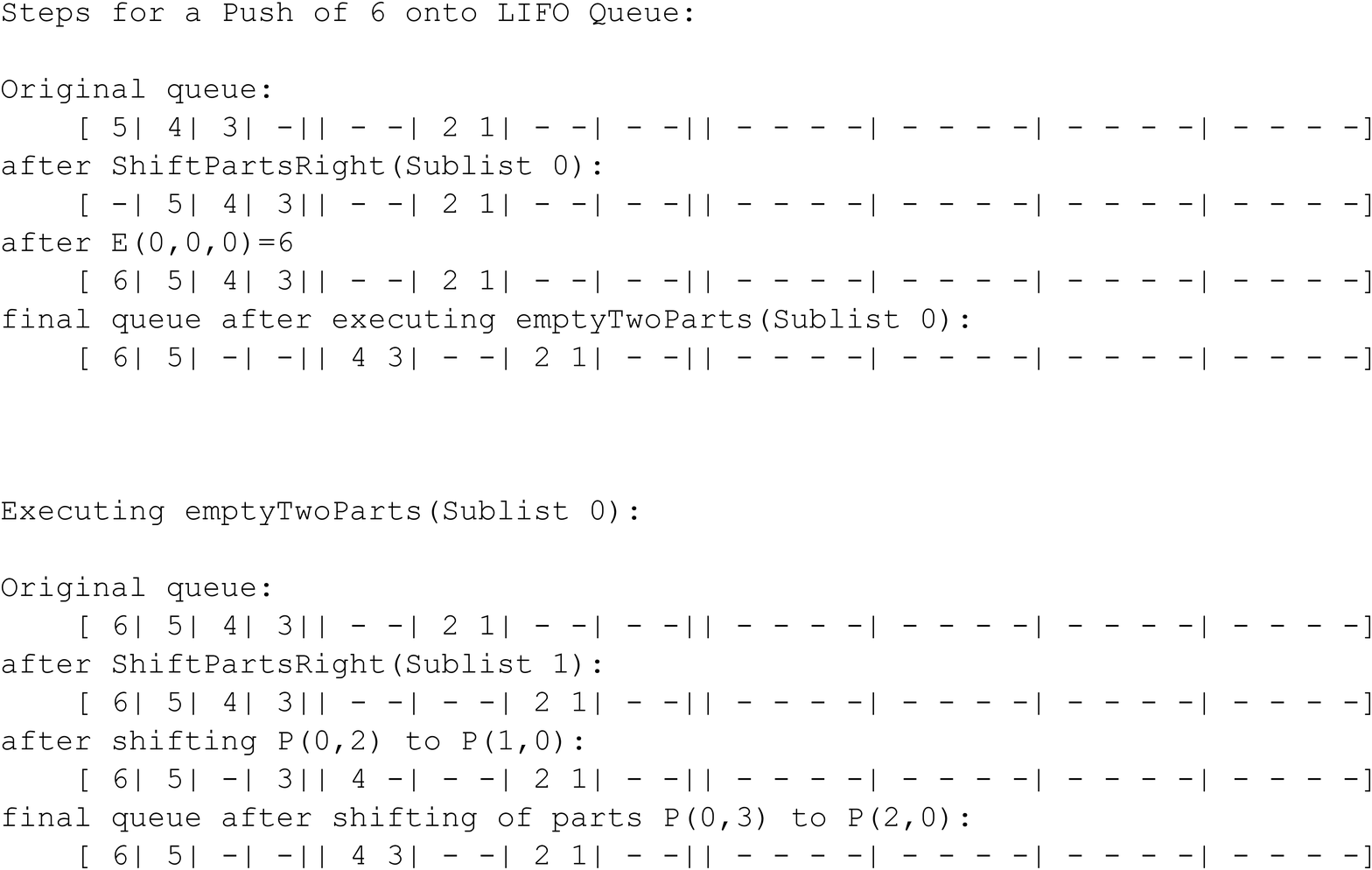}}
	\caption{Steps for pushing an element onto a LIFO queue}
	\label{fig:basic2}
\end{figure*}

More formally, a queue is implemented as an array $A$ that is split into $s$ subarrays (SA) $S_i$ growing exponentially in size with $i$. The total length $n$ of the array is $n:=\sum_{i=0}^{s-1} |S_i|$. Each SA $S_i$ itself is partitioned into $q$ parts $P_{i,0}, P_{i,1},..., P_{i,q-1}$ of equal size $|P_{i,j}|=|S_i|/q$. The size of a part varies for different SAs. We denote the $k$-th element in $P_{i,j}$ by $E_{i,j,k}$. Figure \ref{fig:basic} shows the structure of a queue. 

\begin{algorithm}
	\caption{\textbf{LIFO}} \label{alg:lifo}
	\begin{multicols}{2}
	\begin{algorithmic}[0]	
		\begin{small}
			\STATE \textbf{Initialization(Number of SAs $s$ with $s\geq 1$)}
			\STATE \quad $q:=4$ \COMMENT{number of parts per SA}
			\STATE \quad $E_{i,j,k} := \varnothing$,  \footnotesize{$\forall i \in [0,s-1],j \in [0,q-1],k \in [0,2^i-1]$}
			\STATE \quad $n_{pu} := n_{po} := 0$ \COMMENT{counter for pushes and pops}
			\medskip				
			\STATE\textbf{CmpExEr(b,A,B)}
			\STATE \quad  $A := b\cdot B + (1-b)\cdot A$ \COMMENT{Exchange $A,B$ based on $b$}
			\STATE \quad  $B := (1-b)\cdot B + b\cdot \varnothing$ \COMMENT{Delete $B$ based on $b$}
			\medskip				
			\STATE \textbf{ShiftPartsRight(SA i,doOp)}
			\STATE \quad $empty\&DoOp:= doOp$ if  $E_{i,0,0} \neq \varnothing$ else 0
			\STATE \quad \textbf{for} {SA $j := q-1$ to $1$} \textbf{do} 			
			\STATE \quad \quad $doShift := empty\&DoOp$ if $E_{i,j,0} = \varnothing$ else  $0$
			\STATE \quad \quad \textbf{for} {Element $k=0$ to $|S_i|/q-1$} \textbf{do} 
			\STATE \quad \quad \quad CmpExEr(doShift, $E_{i,j,k}$,$E_{i,j-1,k}$)
			\medskip					
			\STATE \textbf{EmptyTwoParts(SA i)}
			\STATE \quad $isFull := 1$ if $\wedge_{j=0}^{q-1} (E_{i,j,0} = \varnothing)$ else 0
			\STATE \quad ShiftPartsRight(SA i+1,isFull)
			\STATE \quad \textbf{for} {SA $j := q-2$ to $q-1$} \textbf{do} 
			\STATE \quad \quad $o := (j-q+2)\cdot |S_i|/q$ \COMMENT{offset for last 2 parts}
			\STATE \quad \quad \textbf{for} {Element $k:=0$ to $|S_i|/q-1$} \textbf{do} 			
			\STATE \quad \quad \qquad CmpExEr(isFull, $E_{i+1,0,k+o}$,$E_{i,j,k}$)
			\medskip					
			\STATE \textbf{Push(Element x)}
			\STATE \quad $n_{pu} :=$ MoveBetweenSAs($n_{pu}$,EmptyTwoParts)			
			\STATE \quad $b:=1$ if $x\neq \varnothing \wedge E_{i,0,0} \neq \varnothing$ else 0
			\STATE \quad $E_{0,0,0}:=x$ if $x\neq \varnothing$ else $E_{0,0,0}$
			\bigskip
			\STATE \textbf{MoveBetweenSAs(nOps,Operation Op)}					
			\STATE  \quad $mi := 0$ \COMMENT{Find maximal SA to empty/refill}
			\STATE	\quad \textbf{while} $(nOps+1) \mod 2^{mi+1} = 0$  \textbf{do}	
			\STATE \quad \quad $mi := mi+1$
			\STATE \quad \textbf{for} {SA $i:= \min(mi,s-2)$ to 0} \textbf{do} 
			\STATE \quad \quad Apply Operation $Op$ on SA $i$
			\STATE \quad \textbf{return} $(nOps+1) \mod 2^{\max(0,s-2)}$								
			\medskip
			\STATE \textbf{ShiftPartsLeft(SA i)}
			\STATE \quad $full\&DoOp:= 1$ if  $E_{i,0,q-1} \neq \varnothing$ else 0
			\STATE \quad \textbf{for} {SA $j :=$ to $q-2$} \textbf{do} 
			\STATE \quad \quad $doShift := full\&DoOp$ if $E_{i,j,0} \neq \varnothing $ else  $0$
			\STATE \quad \qquad \textbf{for} {Element $k:=0$ to $|S_i|/q-1$} \textbf{do} 
			\STATE \quad \quad \qquad CmpExEr(doShift, $E_{i,j,k}$,$E_{i,j+1,k}$)	
			\medskip					
			\STATE \textbf{RefillTwoParts(SA i)}
			\STATE \quad $isEmpty: = 1$ if $\wedge_{j=0}^{q-1} (E_{i,j,0} \neq \varnothing)$ else 0
			\STATE \quad ShiftPartsLeft(SA i+1)
			\STATE \quad \textbf{for} {SA $j := q-2$ to $q-1$} \textbf{do} 	
			\STATE \quad \quad $o := (j-q+2)\cdot |S_i|/q$	\COMMENT{offset for last 2 parts}
			\STATE \quad \quad \textbf{for} {Element $k:=0$ to $|S_i|/q-1$} \textbf{do} 
			\STATE \quad \quad \quad CmpExEr(doShift, $E_{i,j,k}$,$E_{i+1,0,k+o}$)	
			
			\medskip					
			\STATE \textbf{Pop(doPop)}
			\STATE \quad $result := E_{0,0,0}$ if doPop else $\varnothing$
			\STATE \quad $E_{0,0,0}:=\varnothing$ if doPop else $E_{0,0,0}$
			\STATE \quad $n_{po} :=$ MoveBetweenSAs($n_{po}$,refillTwoParts)			
			\STATE \quad \textbf{return} result
		\end{small} 
	\end{algorithmic}
	\end{multicols}
\end{algorithm}

\subsection{Push, Pop and Shifting}
We explain the shifting procedure shown in Figure \ref{fig:basic} for a sequence of push operations. We always push an element onto the first position in the array $A$ (or pop an element from there). After every modification of the queue, we modify (some) SAs to ensure that there is space for further pushes in the first SA. We shift elements to the right. Shifting is only done on a part level, ie. either we shift all elements of a part or none. We perform frequent shifts to overwrite empty small parts near the beginning of the array and less frequent shifts are conducted for larger parts situated towards the end of the array. We shift parts within a SA but also move parts between SAs, ie. either we merge two parts into one or we split a part into two parts. The subroutines for a push shown in Algorithm \ref{alg:lifo} are discussed next. \smallskip\\ 
\underline{ShiftPartsRight and EmptyTwoParts:} ShiftPartsRight shifts elements from one part to the next part (on the right) within a SA. It avoids overwriting of filled parts by checking if the part to be overwritten is indeed empty. To this end, we only check if the first position of a part is empty. No parts are moved, if the first part of the SA is empty. A parameter indicates whether shifting should take place or not. This is necessary to enable executions of push operations that do not modify the queue. If the parameter is false, ie. zero, then no elements are moved. The order of shifting is from back to front, ie. elements of the second to last part are shifted to the last part (given it is empty), then the third to last part is shifted to the second to last (if empty) and so on.  
EmptyTwoParts empties the last two parts of a SA $i$ by merging them to form the first part of SA $i+1$. It first empties the first part in SA $i+1$ by doing a ShiftPartsRight. Emptying only takes place if all parts of SA $i$ are full and SA $i+1$ is not completely full.  Without this condition for $q>2$ a full part would be (continuously) shifted towards the right for repeated insertions of the empty element $\varnothing$. This would lead to empty SAs followed by (partially) full SAs. As a consequence for pop operations we would have to undo the shifting (or search the entire array).\smallskip  \\
\underline{Push:} A push operation first ensures that the first position of the array is empty. Then, it inserts the pushed element at the first position. A push and its suboperations are illustrated in Figure \ref{fig:basic2}.\smallskip \\
\underline{MoveBetweenSAs:} Restructuring is done after every operation starting from some initial SA (down) to the very first SA in the beginning of the queue. The (index of the) initial SA depends on the number of operations and not the number of actual elements in the queue, which we wish to disguise. Parts of a SA are moved to the next SA, once a SA is full. It might seem reasonable to move all parts of a full SA to the next. However, for alternating pushes and pops this might trigger large performance penalties since parts are continuously moved back and forth between SAs.\\
To disguise the number of elements in the queue (and thus parts), we access all parts in the same deterministic manner for any sequence of pushes of fixed length. Since we allow pushes of a special (empty) element that has no impact on the number of stored elements, the number of operations (as an indicator for the actual number of elements contains) is not exact. We assume that the array grows at a maximal rate, ie. every push is done using a non-empty element. Since we always empty two parts of a SA, we must create space in a SA by moving elements, whenever a sequence of operations could have resulted in the filling of two parts of that SA. For example, every push potentially fills one part in the first SA, since they are of size one. Thus, we would empty the first SA after every second push operation. For SA $i$ with parts of size $2^i$, we would move two parts to the next SA after every $2^{i+1}$ operations. But this approach fails for an arbitrary interleaving of operations pops and pushes of empty and non-empty elements. For example, for the following sequence of pushed elements $1,2,3,\varnothing,4,5$, the algorithm would attempt to empty the first SA after having pushed $1,2$ and again after $1,2,3,\varnothing$. The first SA contains 1,2,3 and misses one element to be full. Thus, the SA would not be emptied and two more elements could be (attempted) to be pushed onto the SA before trying to empty it again, but the SA becomes full after pushing one more element. Therefore, we perform restructuring operations more frequently, ie. for SA $i$ we execute EmptyTwoParts after every $2^i$ operations (rather than after $2^{i+1}$). The last SA that can be emptied is the second to last, ie. the one with index $s-2$. The restructuring is done in Algorithm MoveBetweenSAs which executes for a push operation EmptyTwoParts on all parts as described. It takes as input the counter of the current operations and returns the next value for the counter, which is (usually) the counter incremented by 1. However, once the maximal possible SA has been shifted the operation counter is reset to zero, eg. for $s=5$ the counter is reset after $2^{s-2}=8$ operations. The sequence of maximal SA indexes where parts might be moved to the next SA is a repetition of the sequence 0,1,0,2,0,1,0,3.\smallskip\\
\underline{Pop:} The pop operation (and its subroutines) behave analogously to push but reverse. ShiftPartsLeft shifts parts of a SA within the SA towards the beginning. In contrast to ShiftPartsRight, we do not need a parameter to indicate whether we actually perform the operation or not. ShiftPartsLeft only shifts a part, if the first part is empty. RefillTwoParts moves the first part of SA $i+1$ to the beginning of SA $i$. One full part in SA $i+1$ corresponds to two full parts in SA $i$. As for emptying of parts and right shifts, no non-empty parts are overwritten.

\subsection{Analysis} \label{sec:analysis}
\bth \label{thm:secob}
The LIFO queue is oblivious.
\ethe 
\bpr
According to Definition \ref{def1} we require that memory accesses are independent of the input. (They are allowed to be dependent on the number of operations.) None of the procedures in Algorithm \ref{alg:lifo}  accesses memory cell dependent on an input value, ie. all loop-conditions do not depend on the input and any conditional access to memory cells of the form `cell0:=a if cell1=x else b' can be expressed as multiplications (Section \ref{sec:pre}).
\epr

We analyze push and pop operations with respect to time complexity (Theorem \ref{thm:secp}) and correctness (Theorem \ref{thm:secm}). In the worst case a single operation might take $\Omega(n)$, where $n$ is the maximal length of the queue. We prove that on average, the time is only logarithmic in $n$.

\bth \label{thm:secp}
For the LIFO queue a pop and push operation requires amortized $O(\log n)$ time, ie. $14 q \log(n/q)$ E-Ops and $8q+2$ C-Ops.
\ethe
\noindent The proof uses that two parts of SA $i$ of length $2\cdot 2^i$ are refilled (emptied) after every $2^{i}$ push (pop) operations. Since there are $O(\log n)$ SAs we get time $\sum_{i=0}^{\log n} 2^{i+1}/2^{i} = O(\log n)$.
\bpr
SA $i$ is refilled (emptied) after every $2^{i}$ pop (push) operations. After refilling (emptying) all SAs from index $s-2$ to $0$, ie. after $2^{s-2}$  pop (push) operations, we start over by considering SA 0 only. The average run-time increases up to the point, where SA $s-2$ is considered. Thus, it suffices to compute the average number of operations for a sequence of $2^{s-2}$ pop (push) operations. We analyze pop operations by counting of E-Ops followed by C-Ops.
CmpExEr needs 7 E-OPs (2 additions, 1 subtraction, 4 multiplications).
ShiftsPartsLeft for SA $i$ needs $2^{i}\cdot 7(q-1)$. RefillTwoParts on SA $i$ performs one shift in list $i+1$ and moves one part of it, yielding $2^{i}\cdot 7(q-1) + 2^i\cdot 7q =7\cdot 2^i\cdot q$. Since RefillTwoParts on SA $i$ is called after every $2^{i-1}$ pops, on average refilling of SA $i$ contributes by $7\cdot 2^i\cdot q/2^{i-1}=14q$ E-Ops. By definition we have $n=\sum_{i=0}^{s-1} |S_i|= \sum_{i=0}^{s-1} q\cdot 2^i =  q\cdot (2^{s}-1)$ yielding $s=(\log (n/q))+1$. Summing over all SAs gives $$\sum_{i=0}^{s-2} 14q = 14q(s-1) = 14q\log(n/q)$$
The analysis of C-Ops is analogous. CmpExEr contains zero comparisons. In ShiftPartsLeft we perform one comparison (line 1) and one in each of the $2^i \cdot (q-1)$ iterations. A refill of SA $i$ takes $2q$ comparisons ($q$ to compute isEmpty in RefillTwoParts (line 1) and $q$ within ShiftPartsLeft. Therefore, the number of comparisons becomes  $\sum_{i=0}^{s-2} 2q/2^{i-1}\leq 8q$. Adding two C-Ops due to lines 1-2 in Algorithm Pop completes the proof for pop. The push operation is analyzed in the same manner.
\epr

\ble \label{lem:2states}
Each part $P_{i,j}$ can only be in one of two states: empty (all elements being $\varnothing$) or full (no elements being $\varnothing$).
\ele
\noindent This follows since we modify either all or none of the elements of a part.
\bpr
Initially, all parts are empty. Parts of the first SA can only be full or empty, since they contain at most one element. Parts of SA $i>0$ are only modified due to mergers, splits and shifting. 
Right or left shifting of a part within a SA is done for entire parts. The part overwritten is an exact copy of the part being shifted. The part being shifted becomes empty, ie. all elements are set to the empty element. When all parts of SA $i$ are full, the last two parts of a SA, ie. $P_{q-2,i}$ and $P_{q-1,i}$, each of size $2^i$ are shifted to the next SA, ie. to become the first $P_{0,i+1}$ of size $2^{i+1}$. This part is filled completely. A filled part in SA $i$ (see procedure RefillTwoParts) split into two parts of the same size, yields two full parts in SA $i-1$. 
\epr

\bth \label{thm:secm}
The LIFO queue works correctly.
\ethe

\noindent  We show that no elements are overwritten and no empty elements are returned if the array is non-empty since we refill and empty parts of SAs sufficiently often.
\bpr
In Algorithm \ref{alg:lifo} no parts are overwritten if the first element of a part is non-empty -- see definition and usage of variables $full\&doOp$, $empty\&doOp$ in ShiftPartsLeft/Right; $isFull$, $isEmpty$ in Empty/RefillTwoParts; line 2 of push with $E_{i,0,0} \neq \varnothing$. Since all elements of a part are either the empty element or differ from it (Lemma \ref{lem:2states}), checking the first element suffices to avoid overwriting of non-empty parts.\\
We first show that there is no interleaving of empty and non-empty SAs. Let the $t$-th SA be the largest SA such that at least one part is full. All SAs $i<t$ contain at least one non-empty part. An arbitrary sequence of pushes cannot reduce the number of full parts in a SA below two. This follows since we only empty two parts of a SA if all four parts are full. An arbitrary sequence of pops cannot completely empty a SA except the last, since SA $i$ being of size $q\cdot 2^i$ is refilled with elements from SA $i+1$ after every $2^{i}$ pops (see MoveBetweenSAs in Algorithm \ref{alg:lifo}).\\
Next we show that there is no interleaving of SAs with some non-empty parts and SAs with only full parts. EmptyTwoParts executes on SA $i$ before it executes on SA $j<i$. Upon execution there are two possibilities: Either no or two parts are moved to SA $i+1$. In the first case at most 3 parts are full in SA $i$ and thus, we could insert one more part, in the second case the SA is full and two parts are emptied. Either way, it suffices to empty SA $i$ after two parts in SA $i-1$ (might) have been filled. Since this corresponds to $2^{i-1}$ elements, our choice of calling EmptyTwoParts $i$ after every two $2^{i-1}$ operations suffices (see MoveBetweenSAs in Algorithm \ref{alg:lifo}). Therefore, not all parts of a SA can be full, if there is space in a larger SA. For refilling parts an analogous argument applies.
\epr

\section{FIFO} \label{sec:fifo}
\begin{figure*}[!htp]
	\centering
	\centerline{\includegraphics[width=\linewidth]{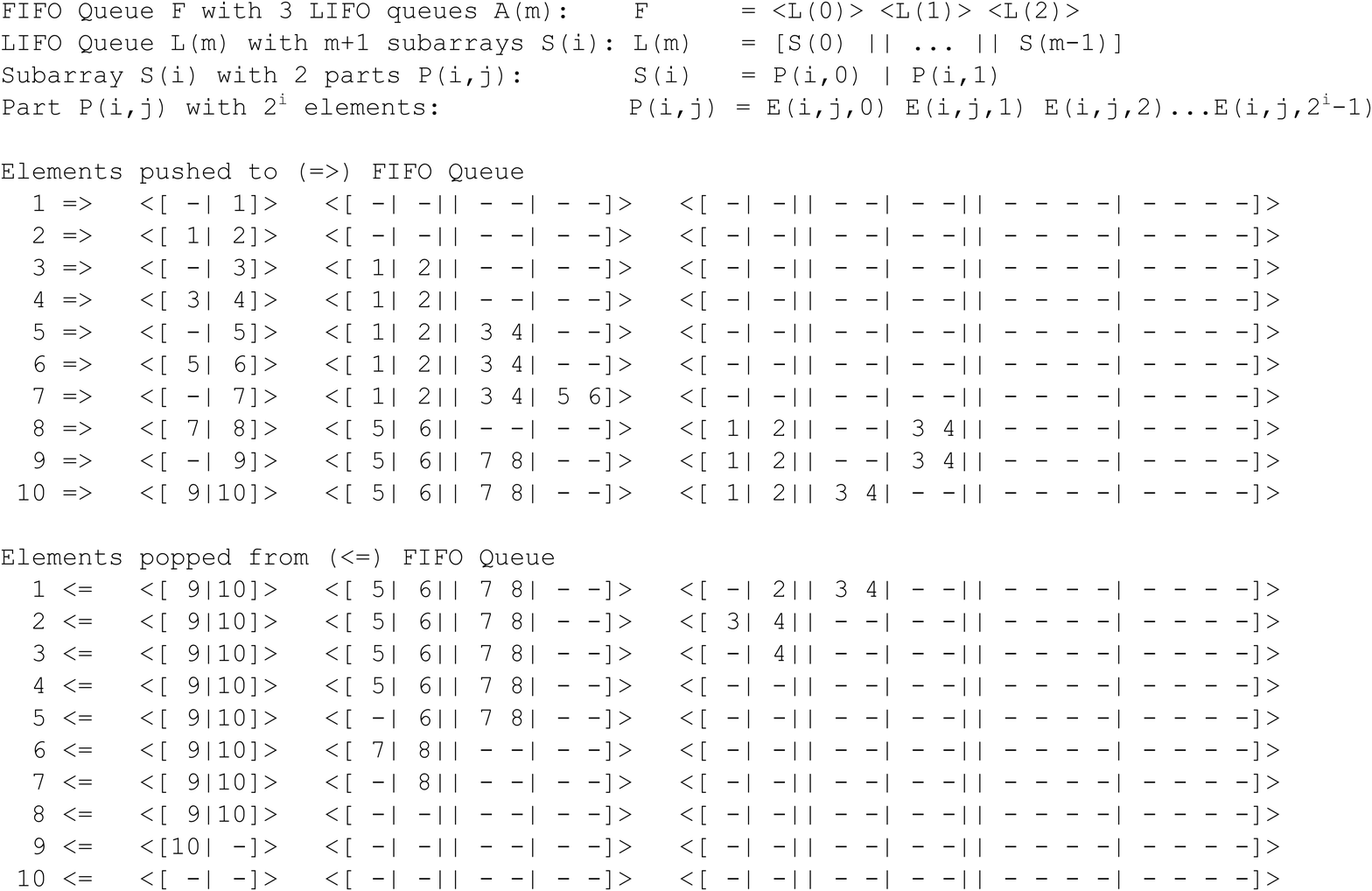}}
	\caption{Sequence of pushes and pops onto a FIFO queue}
	\label{fig:basicfifo}
\end{figure*}

A First-In-First-Out (FIFO) queue needs fast access to the first and the last element. We use an array of LIFO queue variants of increasing lengths, ie. each SA of the FIFO queue is itself a LIFO queue. Each LIFO queue stores elements in `reverse' order, meaning the first element to be popped in the LIFO queue is the oldest element the LIFO queue contains. In this way we can efficiently access the oldest element of each LIFO queue. The array structure is visualized in Figure \ref{fig:basicfifo}. Each LIFO queue matches on SA.

For a pop operation the LIFO queue with largest index that is non-empty is identified. Then an element is popped from that queue. To make the algorithm oblivious we execute a pop operation on every LIFO queue within the FIFO queue. We start from the back and pop an element from each LIFO queue, ie. SA, until the first non-empty LIFO queue has been identified. For the remaining queues we execute pops using a parameter to indicate that, in fact, no element should be popped. The key point is that indepedent of the value of the parameter the same memory cells are accessed.

\underline{PopperQueue}: A LIFO queue offers more functionality than is needed, since we do not push elements in the front but only pop them except for the first queue, which is just a single element. Opposed to a LIFO queue, we can therefore refill a SA completely. We reduce the number of parts from four to two. Using more parts per SA is slower since we must shift the same elements multiple times rather than moving them less often in bigger chunks, ie. larger SAs. We can reuse most LIFO procedures (Algorithm \ref{alg:lifo}) without modification, ie. ShiftsPartLeft, RefillTwoParts and Pop. We call this LIFO variant ``PopperQueue''. It is a special case of the LIFO queue from Section \ref{sec:lifo}. It has the same (asymptotic) properties, but it is roughly a factor of two faster, since it uses less parts and therefore requires less shifts within a SA, ie. compare Theorem \ref{thm:secp} for $q=4$ (LIFO) and $q=2$ (PopperQueue).

\smallskip

Due to the more involved array organization of a FIFO queue, the emptying of parts and refilling of parts needs careful attention. It is not possible to concatenate two parts to get a larger part without extra processing, ie. two arrays (of PopperQueues) placed after each other generally do not yield an array representing a larger PopperQueue with a valid structure. The concatenation could give partially filled parts. For example, assume that there are two queues with one SA and two parts, eg. $[1|-]$ and $[-|4]$, naive concatenation yields $[1|-|| - 4|- -]$ having the partially filled part $|- 4|$. Furthermore, we have to ensure a correct ordering of the elements within LIFO queues when moving elements between them.

If one last part of the PopperQueue stored in SA $i$ is full, we move $2^i$ elements from queue $i$ to the very last part of queue $i+1$. We pop one element after the other from queue $i$ and put it directly into the last part of queue $i+1$, ie. the element of the $j$-th pop is put at the $j$-th position of the last part. At this point the whole queue $i+1$ (except the last part that was just inserted) might be empty which would cause subsequent calls of pop on queue $i+1$ to fail. Therefore, we attempt to shift elements from the last part of the last SA of LIFO queue $i+1$ consisting of newly inserted elements up to the first SA of queue $i+1$.\\
The push operation for the FIFO queue appends elements to the end of the very first LIFO queue. Since it is of length two, we shift the second element of it to the left and then set the second position to the newly inserted element.

\bco \label{thm:secm2}
For the FIFO queue a pop operation requires $O(\log^2 n)$ and a push $O(\log n)$ time on average.
\eco
\bpr
For a pop of the FIFO queue we do a pop for each of the PopperQueues giving $\sum_{i=0}^{s} O(i) = O(s^2) = O(\log^2 n)$.
For a push we move blocks of size $2^i$ from SA $i$, ie. PopperQueue $i$, to SA $i+1$ after every $2^i$ operations, which needs time linear in the queue length. Summation gives $\sum_{i=0}^{s} O(2^i/2^i) = O(s) = O(\log n)$.
\epr

\subsection{Fast FIFO (and Double-Ended Queues)} \label{sec:fast}
FIFO queues are often used as buffers to distribute peak loads across a longer timespan. Commonly, a producer pushes elements onto the queue continuously (as a stream), while a consumer repeatedly takes an element and processes it. Buffering always introduces some delay in processing. Thus, usually an additional delay is tolerable. A pop on the fast FIFO queue only returns an element given the queue has been filled partially, ie. it is at least half full.\\
Our FIFO queue that has only amortized $O(\log n)$ overhead rather than $O(\log^2 n)$. The idea is to use two queues ``back to back'': one for popping and one for pushing. The two queues share the last part, ie. both treat this part as belonging to them. Thus, elements are pushed onto one of the queues and are continuously shifted to the right with newly inserted elements until they reach the queue for popping. A pop only returns an element after its last part of the last SA (shared with the pushing queue) has been filled. 
The same ideas also apply to double-ended queues.\\ 

\noindent For the Fast FIFO Queue (B2B Queue) the time complexity of a push and pop matches the corresponding operations for the LIFO queue.
\bco \label{thm:fifo}
For the B2B-FIFO Queue a pop and push operation require O($\log n$).
\eco

\section{Double-Ended Queue}\label{sec:array}
A double-ended queue supports popping elements at the head and tail as well as prepending elements at the beginning and appending them at the end. We combine ideas for LIFO and FIFO queues. We use an array of queues (as for FIFO queues) to address the need to push elements to the head of the array and pop them from the tail. Since elements can also be pushed at the back, we use LIFO queues, ie. SA $i$ of the double-ended queue is given by a LIFO queue with $i+1$ SAs (rather than PopperQueues). Pushing to the back requires identifying the last non-empty SA, ie. queue, as for popping from the back in the FIFO queue. However, we can only push the element onto the queue, if it is non-full, otherwise we push it onto the next queue.
Popping elements from the front might trigger refilling of SAs. In turn, we have to move the newest elements of one SA to another. Identifying the newest elements of a LIFO queue (with elements sorted by age, ie. ascending insertion order) is cumbersome, since there is only efficient access to the oldest element. To reverse order, we remove all elements from the array (using a sequence of pops) and insert them into a temporary LIFO queue. This yields a queue sorted by newest to oldest elements. Then we move elements by popping them from the temporary queue to the queue to refill, ie. for queue $i$ we move $2^{i+1}$ elements. The remaining elements are pushed back onto the emptied queue (used to create the temporary LIFO queue).

\bth \label{thm:doe}
Any operation on the double-ended queue has amortized time $O(\log^2 n)$.
\ethe

\noindent Operations are similar to the LIFO queue, except for refilling and emptying that needs an additional logarithmic factor due to the popping and pushing of elements rather than direct access in O(1).
\bpr
Pushing and popping to the front works the same as for LIFO queues except for the refilling and emptying of full SA. We require an additional logarithmic factor, since we cannot just copy elements of one SA, ie. queue, to another but we first pop them from the LIFO queue onto a temporary queue. More precisely, each element access using a pop requires amortized O($\log n$) as shown in Theorem \ref{thm:secp} rather than O(1).
Pushing and popping to the back requires executing a constant number of push and pop operations for all parts constituting LIFO queues. Since we have O($\log n$) queues and each operation on a LIFO queue requires O($\log n$) (see Theorem \ref{thm:secp}), a single push and pop operation requires $O(\log^2 n)$.
\epr

\section{Double-Ended Priority Queue} \label{sec:prio}
In this scenario, each data item has a priority. A double-ended priority queue can return either the data element with the smallest or largest priority. The queue structure is the same as for double-ended queue. We ensure that each SA, ie. LIFO queue, contains elements sorted in descending priority. When moving elements from one queue to another, ie. to empty a full queue or refill a queue, we first create one single sorted array containing all elements from both queues and then refill the smaller queue up to half of its capacity with the elements of smallest priority and put the other elements in the larger queue. The sorting can be done by merging both arrays.


Popping the element of minimum priority requires finding the smallest element in SA 0. Popping the element of maximum priority requires checking all parts, since we do not know which parts contain elements and which do not as well as which part contains the element with largest priority. More precisely, we first (peek) all parts and find the element and part with the maximum element. After that we perform a pop on the (first) queue containing the maximum element. This is done by executing a pop for all parts. The parameter of the pop operation, determining whether the operation indeed removes an element from the queue, must be set accordingly, ie. it is $\varnothing$ for all but the queue containing the maximum element. 



The restructuring is somewhat more involved. Upon a push that requires restructuring, eg. either refilling or emptying queue $i$ we first create one sorted array in increasing order by merging both queues as done for ordinary mergesort (see also Section \ref{sec:msort}). We then refill SA $i$ until it is half full with the smallest elements (in reversed order) and insert the remaining to the next SA (in reversed order).

\bth \label{thm:dprio}
Any operation on the double-ended priority queue has amortized time $O(\log^2 n)$.
\ethe
\bpr
We discuss time followed by correctness.
Pushing an element to the front (or popping the element of minimum priority) works the same as for LIFO queues except for the emptying and refilling of full SA. We require an additional logarithmic factor to move elements from queue $i$ to queue $i+1$ (or the other way around), we create a temporary queue by repeatedly popping the element of maximum priority and adding it to the temporary queue. Each element access using a pop requires amortized O($\log n$) as shown in Theorem \ref{thm:secp} rather than O(1). Moving the elements from the temporary queue onto the (new) queues $i$ and $i+1$ has the same asymptotic time complexity. Therefore, push to the front need $O(\log^2 n)$ time.
Popping the maximum priority element requires executing a pop operation for all LIFO queues (plus restructuring). Since we have O($\log n$) queues and each operation on a LIFO queue requires O($\log n$) (see Theorem \ref{thm:secp}), a single push and pop operation requires $O(\log^2 n)$.
Popping the maximum priority element requires executing a pop operation for all LIFO queues (plus restructuring). This requires $O(\log^2 n)$.\\
Correctness of a pop of maximum priority follows, since we maintain all queues in descending order of priority.  Thus, the element of maximum  priority is the first element in one of the queues. Since we consider the first elements of all queues and return the one of maximum priority, correctness follows. For the minimum we only investigate the first queue. Since upon every restructuring operation on queue $i$ we keep the smallest half of both queues $i$ and $i+1$ in queue $i$, it holds that after a restructuring all elements in SA $i$ are smaller than any element in SA $i+1 > i$. Using induction, we have that the smallest element is in SA 0.
\epr

\section{Oblivious Mergesort} \label{sec:msort}
Our oblivious mergesort algorithm (O-Mergesort) divides an unsorted array (or list) into SAs of one element. It repeatedly merges two arrays of equal length to obtain a new sorted array of double the length until there is only one array remaining. This array is sorted. To make the sorting procedure oblivious requires a queue that supports a conditional pop, ie. we pop the element of the array if it is smaller than another element. For short arrays (of length 1), we use a naive sort. Otherwise, two PopperQueues are merged by repeatedly comparing the first element of each queue $A$ and $B$ and appending the smaller one to the result array $C$. Note, that since $A$ and $B$ are sorted the element put into $C$ is the smallest element in both arrays. We pop an element from the array which element we just appended to $C$ -- see Algorithm O-Merge \ref{alg:merge}.  

\begin{algorithm}[!htp]
	\caption{\textbf{O-Merge}} \label{alg:merge}
	\begin{algorithmic}[1]
				\REQUIRE{ Sorted PopperQueue $A$ and $B$ of length $l$}
				\ENSURE{ Merged LIFO Queue $C$ } 
				
		\begin{small}
			\IF{$l=1$}
				\STATE $b:= 1$ if $A[0]\leq B[0]$ else 0
				\STATE $C[0]:=A[0]\cdot b + B[0]\cdot (1-b)$
				\STATE $C[1]:=B[0]\cdot b + A[0]\cdot (1-b)$			
			\ELSE
			\STATE $eleA:=A.pop(1)$ \COMMENT{Returns smallest element in $A$}
			\STATE $eleB:=B.pop(1)$
			\FOR{$k=0 \textbf{ to } 2\cdot l-2$}
			\STATE \COMMENT{Set $C[k]$ to the smallest element in $A$ union $B$ and remove the element}
			\STATE $b:= 1$ if $eleA\leq eleB$ else $0$
			\STATE $C[k]:=eleA\cdot b + (1-b)\cdot eleB$					
			\STATE $eleA := A.pop(b)\cdot b +(1-b)\cdot eleA$ 
			\STATE $eleB := A.pop(1-b)\cdot (1-b) +b\cdot eleB$			
			\ENDFOR
			\STATE $C[2\cdot l-1]:=eleA\cdot b + (1-b)\cdot eleB$					
			\ENDIF			
		\end{small}
	\end{algorithmic}
	\small
\end{algorithm}

\bth \label{thm:msort}
Sorting of an array of $n$ elements requires at most  $85 n \log n$ C-Ops and a total of $n\cdot (3+ 560(\log n-1) +28\log^2 n)$ E-Ops.
\ethe
\bpr
The merger of two arrays of size $l$ each requires $4l$ pop operations, each requiring $18$ comparisons using Theorem \ref{thm:secp} with $q=2$. Additionally, we need one more comparison per iteration. This gives a total of $85l$ S-Ops for merging two arrays.
In total we get the following bound $\sum_{j=0}^{\log n - 1}   2^{\log n - j} \cdot 85 \cdot 2^j \leq 85 n\log n$ S-Ops.

The naive sort of two arrays of size one comparing the two elements requires 5 E-Ops. In total there are $n/2$ queues of length 1 yielding a total of $5n/2$.
The merger of two arrays of size $l>1$ requires each $4l$ pop operations, each requiring $28 (\log(l/2)+1)=28\log l$ E-Ops.\footnote{We have $\log (l/2)+1$ rather than $\log (l/2)$ using Theorem \ref{thm:secp} with $q=2$ since we merge arrays of length $l=2^x$ but we only support mergers of arrays of length $2^{y}-1$, thus we need $y=x+1=\log (l/2) +1$}, giving a total of $112 l\log l$. Additionally, we need 5 E-Ops for each of the $2l$ operations, giving a total of $10 l$ E-Ops. Overall we get $l(10+112 \log l)$. 
Overall, we get 
\small{
	\[ \begin{aligned}
	&5n/2+ \sum_{j=1}^{\log n - 1}   2^{\log n - j-1} \cdot 112\cdot 2^j\cdot  (\log (2^j)+10)\\
	&= 5n/2+\sum_{j=1}^{\log n - 1}   n \cdot 56\cdot  (\log (2^j)+10)\\
	&\leq 3n+56 n (\log^2 n/2+10(\log n-1))\\ 
	&= n\cdot (3+ 560(\log n-1) +28\log^2 n)
	\end{aligned}   \]}
\epr

\noindent The analysis uses that we merge $\frac{n}{2^{i}}$ arrays of length $2^{i}$ and Theorem \ref{thm:secp} to bound the time two merge two arrays. We improve on \ci{goo14} by a factor of more than 1000 in terms of the number of comparisons, ie. C-Ops. Comparisons are often used to analyze sorting algorithms, since typically the total operations involved is proportional to the number of comparisons. In our case, this does not necessarily hold, since we only require one comparison for shifting a large number of elements. Therefore, the costs for shifting might dominate the costs for comparisons. To ensure a fair and objective comparison among algorithms we also analyzed the number of other operations, ie. E-Ops, since they are the dominant factor in our algorithm. With respect to the total number of operations O-Mergesort is asymptotically worse by a factor $\log n$. However, due to the extremely large constants used in the state-of-the-art\ci{goo14} we use less operations for all practically relevant scenarios, ie. for arrays of length up to roughly $2^{5300}$. For illustration, when sorting 10 billion elements we need more than 100x less E-Ops. Furthermore, E-Ops (or XORs, ANDs) are generally less complex than comparisons, therefore in practice the speed-up might be even larger.

\section{Quicksort} 
Our oblivious quicksort algorithm (O-Quicksort) is a comparison-based divide and conquer algorithm. Small arrays of size at most $4\log^2 n$ are sorted using O-Mergesort. Larger arrays are recursively split into two smaller (sub)arrays. An array is split using a pivot element. All elements less or equal to the pivot are put in one array and all larger elements in the other array. Ideally, both arrays are of the same size. However, naive splitting likely leads to badly balanced arrays leading to $O(n^2)$ run-time since an oblivious algorithm must treat both parts as potentially large. However, when choosing the median as pivot, it is possible to ensure that both arrays are of equal size. We compute an approximate median for all elements (Section \ref{sec:ranp}). Unfortunately, choosing an approximate median still leaves some uncertainty with respect to the exact array lengths after the splitting. Therefore, in the partition process (Section \ref{sec:ranp}), rather than swapping elements within one array, we create two arrays of fixed length, one for elements larger than the pivot and one for all other elements. Since the length of each of the two arrays must be fixed using conservative upper bounds, their sum of lengths exceeds the length of the array to be split. To get a single sorted array requires a special reunification of both arrays (Section \ref{sec:squi}).  For simplicity, we assume that all elements to be sorted are distinct. This assumption is removed in Section \ref{sec:dup}. 

\subsection{Random Pivot Choice and Partition} \label{sec:ranp}
Algorithm RandomPivot chooses several elements uniformly at random, sorts them and picks the median. By choosing the median of a sufficiently large sample of elements we ensure that the chances of a split resulting in very unbalanced arrays is small. We pick a fixed number of samples $n_p$, sort them, eg. using the O-MergeSort algorithm, and then pick the middle element $l/2$ of the sorted array of length $l$ as pivot.

\begin{algorithm}[!htp]
	\caption{\textbf{RandomPivot}}\label{alg:ran}

	\begin{algorithmic}[1]
				\REQUIRE{ Array $A$ of length $l$, number of samples $n_p$}
				\ENSURE{ Pivot $p$ } 
		\begin{small}
			\STATE $P :=$ Set of $n_p$ elements of $A$ chosen uniformly at random
			\STATE $SP :=$ Sorted samples $P$ \COMMENT{eg. using O-MergeSort}
			\STATE $p:= SP[l/2]$ \COMMENT{Choose middle element (= Median) as pivot}
		\end{small}
	\end{algorithmic}
	\small
\end{algorithm}

For the partitioning the entire array is split into two arrays, one with all elements being smaller than the pivot and one with all elements being larger. The two arrays are given by two LIFO queues. We push elements that are smaller than the approximated median on one of the queues and the larger elements on the other queue. We discuss the case of duplicates in Section \ref{sec:dup}.
\bth \label{thm:piv}
Algorithm RandomPivot returns a pivot $p$ such that at least a fraction $c_f= \frac{1}{2} \cdot \dfrac{1}{1+\sqrt{13c(\log n)/n_p}}\geq 1/4$ of elements of an array of length $n$ are larger than $p$ and the same fraction is smaller than $p$ with probability $1-1/n^c$ for $n_p\geq 13\cdot c\cdot \log n$. 
\ethe
\noindent We obtain tail estimates using carefully applied Chernoff bounds.
 \bth[Chernoff Bound] \label{thm:Che}
 The probability that the number $X$ of occurred independent events $X_i \in \{0,1\}$, i.e. $X:=\sum_i X_i$, is not in $[(1-c_0)\mathbb{E}[X],(1+c_1)\mathbb{E}[X]]$ with $c_0\in]0,1]$ and $c_1\in]0,1[$ can be bounded by $p(X \leq (1-c_0)\mathbb{E}[X] \vee  X \geq (1+c_1)\mathbb{E}[X]) < 2e^{-\mathbb{E}[X]\cdot \min(c_0,c_1)^2/3}.$
 \ethe

\noindent Proof of Theorem  \ref{thm:piv}:
\bpr
The value of $c_f$ is minimized, when $n_p$ is smallest. Thus, the bound $c_f\geq 1/4$ follows from substitution of the lower bound, ie. $n_p=13\cdot c\cdot \log n$, into $c_f= \frac{1}{2} \cdot \dfrac{1}{1+\sqrt{13c(\log n)/n_p}} = 1/2\cdot \frac{1}{1+\sqrt{1}}=1/4$.
The theorem holds if the pivot does not stem from the $c_f\cdot n$ smallest or largest elements. If we pick less than $c_f\cdot n_p <n_p/2$  elements $S \subseteq A$ from the $c_f\cdot n$ smallest and less than $c_f\cdot n_p < n_p/2$ elements $L\subseteq A$ from the $c_f\cdot n$ largest elements this will be the case. The reason being that the pivot $p$ is the element at position $n_p/2$  in the sorted sequence and, thus, it will not be from the set of $c_f\cdot n$ smallest or largest elements.  We expect to pick $c_f \cdot n_p$ elements $S$ out of the $c_f\cdot n$ smallest elements (and analogously for the largest), ie. $E[|S|]=c_f\cdot n_p$. We seek the smallest factor $f>1$ such that when exceeding the expectation by factor $f$ the pivot is not chosen correctly. We have $f\cdot c_f\cdot n_p = n_p/2$, if $f=1/(2\cdot c_f)$. The probability that the expectation is exceeded by a factor $f>1$ or more is given using a Chernoff bound (see Theorem \ref{thm:Che}) by 
\begin{small}
	\[ \begin{aligned}
	& prob(|S|>f\cdot E[S])  < 1/2^{(f-1)^2/3 \cdot c_f \cdot n_p} 
	\leq 1/2^{1/12\cdot (f-1)^2 \cdot n_p} \text{ (Using $c_f\geq 1/4$)}\\
	& = 1/2^{1/12\cdot (1/(2\cdot c_f)-1)^2 \cdot n_p} \text{ (Using $f=1/(2\cdot c_f)$)}\\
	& = 1/2^{1/12\cdot ((1+\sqrt{13c\log n/n_p})-1)^2\cdot n_p}
	= 1/2^{13/12\cdot c\log n} = 1/n^{13/12\cdot c}
	\end{aligned}   \]
\end{small}
In the same manner we can compute $prob(|L|>f\cdot E[L])$. Therefore the probability of both events becomes for $n$ sufficiently large:
\small{
	\[ \begin{aligned}
	prob(\big(|L|\leq f\cdot E[L]\big) \wedge \big(|S|\leq f\cdot E[S]\big) ) 
	\geq 1- \big(prob(|L|>f\cdot E[L])+ prob(|S|>f\cdot E[S]\big) \geq 1-1/n^c  
	\end{aligned}   \]}
\epr

\begin{figure}[htp!]	
	\centering{
		\includegraphics[width=0.75\linewidth]{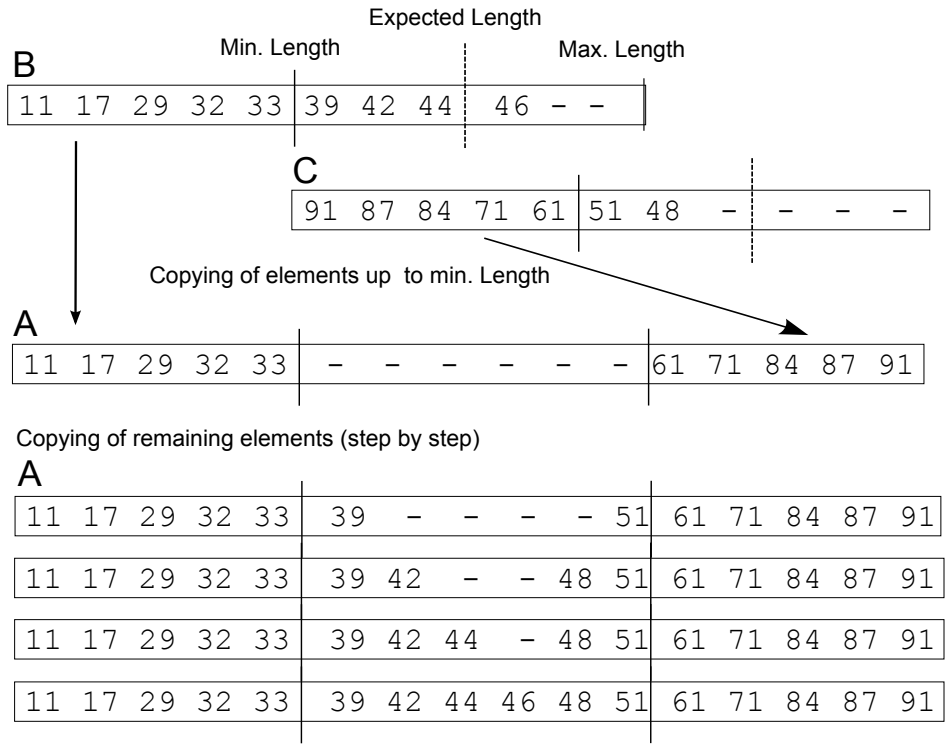}}
	\caption{Merger of two subarrays within O-Quicksort}
	\label{fig:merger}
\end{figure}

\begin{algorithm}[!ht]
	\caption{O-Quicksort} \label{alg:squi}

	\begin{algorithmic}[1]
			\REQUIRE{Array $A$ of length $l$, Sort ascending: $asc$}
			\ENSURE{ Sorted array $A$ } 
		\begin{small}
			\STATE $\epsilon:=\sqrt{13\cdot c\cdot (\log n)/l}$
			\IF{$l\geq 4 \log^2 n$}
			\STATE $B,C:=Partition(A)$
			\STATE O-Quicksort($B,l/2\cdot (1+\epsilon),True)$)
			\STATE O-Quicksort($C,l/2\cdot (1+\epsilon),False)$)
			\FOR{$k=0 \textbf{ to } l/2\cdot (1-2\epsilon)$}
			\STATE $B[k]:=B[k]$ \COMMENT{Copy elements from $B$ to $A$}
			\STATE $B[l-k]:=C[k]$ \COMMENT{Copy elements from $B$ to $A$}
			\ENDFOR
			\FOR{$k=l/2\cdot (1-2\epsilon) \textbf{ to } l/2\cdot (1+\epsilon)$}
			
			\STATE \textbf{if }{$B[k]\neq v_{\varnothing}$}\textbf{ then } $B[k]:=B[k]$ \COMMENT{Copy elements from $B$ to $A$}
			\STATE \textbf{if }{$C[k]\neq v_{\varnothing}$}\textbf{ then } $B[l-k]:=C[k]$ \COMMENT{Copy elements from $C$ to $A$}
			\ENDFOR
			\ELSE
			\STATE $B$:= Sort using O-MergeSort or other alg. either ascending or desc. depending on $asc$
			\ENDIF
		\end{small}
	\end{algorithmic}
	\small
\end{algorithm}

\subsection{Sorting and End-to-End Array Merger} \label{sec:squi} 
So far we have obtained well-balanced, but unsorted SAs. Since we do not have access to their exact lengths we use a conservative bound on their lengths given by the analysis of the partition process. O-Quicksort recurses on two separate arrays stemming from the partitioning. We sort the array of smaller elements $B$ than the pivot in ascending order and the array of larger elements $C$ in descending order. At the end, both arrays must be merged to get a single final array of sorted elements. This requires some care since we do not know their exact lengths. Due to the partitioning process we can bound the minimum length of $B$ and $C$ to be $l/2\cdot (1+\epsilon)$ with $\epsilon:=\sqrt{13\cdot c\cdot (\log n)/l}$. We copy the elements (up to the guaranteed minimum length bound) to the final array, so that these elements appear sorted. This means we fill the final array with $B$ from the left end towards the right and with $C$ from the right end. The entire process is illustrated in Figure \ref{fig:merger}. The remaining elements are handled in the same fashion, but before setting an array element in $A$ to be an element from $B$ or $C$, we check whether the element in $A$ is still empty. 

\bth \label{thm:secq}
O-Quicksort needs $O(n \log n)$ C-Ops and E-Ops. It produces a correct sorting with probability $1-1/n^c$ for an arbitrary constant $c$.
\ethe
\noindent The recurrences are somewhat involved, since the lengths of both arrays used for recursion exceeds the original length of the array being split. We conduct a staged analysis to obtain (asymptotically tight) bounds.
\bpr
The complexity $T(n)$ of O-Quicksort in terms of comparisons can be stated using a recurrence. For one call (ignoring recursion) to an array of length $l>4\log^2 n$ we have that the complexity is given by partitioning the array being $O(l)$ plus the reunification of both sorted arrays, ie. the copying of elements being also $O(l)$. Thus, we get a total of $O(l)=c_0\cdot l$ for some constant $c_0$. We obtain the following recurrence for an array of length $l$ using $\epsilon=\sqrt{a/l}$ with $a:=13\cdot c \cdot (\log n)$: 

\begin{small}
	\[ \begin{aligned}
	\text{First call: } & T(l)= 2T(l/2\cdot (1+\sqrt{a /l}))+ c_0\cdot l\\
	\smallskip
	\text{Second call: } &T(l/2\cdot (1+\sqrt{a /l}))  \leq 2T(l/4\cdot (1+\sqrt{a /l})\cdot (1+\sqrt{a /(l/2)})) + c_0\cdot l/2\cdot (1+\sqrt{a /l}) \\
	& \leq 2T(l/4\cdot \big(1+\sqrt{a /(l/2)}\big)^2)  + c_0\cdot l/2\cdot (1+\sqrt{a /l})\\	
	\end{aligned}   \]
\end{small}
\begin{small}
	\begin{align}
	\text{Third call: } & T(l/4\cdot (1+\sqrt{a /(l/2)})^2)  \leq 2T(l/8\cdot (1+\sqrt{a /(l/4)})^3) + c_0\cdot n/8\cdot (1+\sqrt{a /(l/4)})^3 \nonumber\\
	\smallskip
	\text{$i$-th call: } &T(l/2^i\cdot \big(1+\sqrt{a /(l/2^{i-1})}\big)^i) \leq 2T(l/2^{i+1}\cdot \big(1+\sqrt{a /(l/2^i)}\big)^{i+1}) + c_0\cdot l/2^i\cdot \big(1+\sqrt{a /(l/2^i)}\big)^{i+1} \label{eq:rec} 
	\end{align}  
\end{small}

Assume we start splitting the entire array $A$ with $l=n$. The total number of operations (C-Ops and E-Ops) at recursion depth $i$ is given by the additive term in Equation (\ref{eq:rec}) multiplied by the number of calls to O-Quicksort being $2^i$, ie.
\begin{small}
	\[ \begin{aligned}
	&2^i c_0\cdot n/2^i\cdot \big(1+\sqrt{a / (n/2^i)}\big)^{i+1} =  c_0\cdot n\cdot \big(1+\sqrt{a /(n/2^i)}\big)^{i+1}
	\end{aligned}   \]
\end{small}
The total operations for the first $r:=\log n - 8 \log \log n$ recursions is given by:
\begin{small}
	\[ \begin{aligned}
	&\sum_{i=0}^{r-1} c_0\cdot n\cdot \big(1+\sqrt{a /(n/2^i)}\big)^{i+1} 
	\leq c_0\cdot n \cdot \sum_{i=0}^{r-1}   \big(1+\sqrt{a /(\log n)^8}\big)^{\log n}\leq c_0\cdot n \cdot \log n
	\end{aligned}   \]
\end{small}
After $r$ recursions the size of the input sequence for the recursive calls is at most $n/2^{r}\cdot (1+\sqrt{a /(n/2^{r-1})})^{r} \leq 2\cdot \log^8 n$ (for $n$ sufficiently large).
For another $6 \log \log n$ recursions on an array of length $2\log^8 n$ the number of operations is bounded by:
\begin{small}
	\[ \begin{aligned}
	&c_0\cdot 2 \log^8 n \cdot \sum_{i=0}^{6 \log \log n-1}   (1+\sqrt{\dfrac{4\log n}{(\log n)^2}})^{6\log \log n} = c_0\cdot 2 \log^8 n \cdot \sum_{i=0}^{6 \log \log n-1}   (1+\dfrac{2}{\sqrt{\log n}})^{6\log \log n}  \leq 4c_0\log^8 n
	\end{aligned}   \]
\end{small}
The size of the remaining arrays is $4\log ^2 n$ using the same derivation as above using $r$ recursions.
To sort such an array using O-MergeSort requires $O(\log^2 n \log \log n)$ C-Ops and $O(\log^2 n (\log \log n)^2)$ E-Ops (see Theorem \ref{thm:msort}).
There are $2^{\log n - 2\log \log n} = n/\log^2 n$ such arrays, giving a total of $O(n \log \log n)$ C-Ops and $O(n (\log \log n)^2)$ E-Ops.
To obtain a correctly sorted queue all executions of RandomPivot must be successful. We perform at most $\log n - 2\log\log n$ recursions. Thus, in total there are at most $n$ calls to RandomPivot, each succeeding with probability at least $1-1/n^{c'}$ for an arbitrary constant $c'$. The probability that all succeed is at least $(1-1/n^{c'})^{n}\geq 1-1/n^{c'-2}$. Choosing $c'=c+2$ concludes the proof.
\epr

\subsection{Equal or Duplicate Elements} \label{sec:dup}
So far we focused on arrays of distinct elements. For non-distinct elements our algorithm can fail to compute balanced arrays in case the chosen median is not unique. In the most extreme case all elements are the same and the split would result in one empty array and one array being the same as the array to be split. Elements can always be made distinct by appending a unique number at the end, eg. by appending a counter to a array of elements $(0,0,0)$, we get $(00,01,02)$.  We assign elements that are equal to the pivot $p$ to both arrays such that their lengths maintain balanced. In a first phase we create two arrays $B,C$ and maintain a counter $l_p$ for the elements equal to $p$ by distinguishing three cases for an element $x$ that is compared to the pivot $p$, ie. $x<p$, $x>p$ and $x=p$. In the first case, we assign $x$ to array $B$ and increment the length counter of $l_B$. In the second case we assign $x$ to $C$ and increment the length counter $l_C$ of $C$. In the third case, we increment just the counter $l_p$. In the second phase we distribute $l_p$ copies of $p$ to the arrays $B$ and $C$ such that their difference in length is as small as possible. We perform $l$ iterations, where $l$ is the number of elements in the array to be partitioned, ie. $A$. In each iteration we subtract one from $l_p$. If $l_p$ is zero the arrays remain the same. Otherwise, if the lengths of $l_B$ is less than $l_C$, we append a copy of the pivot to $B$ and increment the length counter $l_B$ otherwise we do the same for $C$. The (asymptotic) complexity remains the same.

\section{Applications} \label{sec:app}
\subsection{Confidential Expressions} \label{sec:coFi}
Confidential expressions, hiding data as well as operations on the data are a rather straight forward application of our LIFO queue with conditional operations as well as basic operations, e.g. addition and multiplication, from secure computing schemes such as fully homomorphic encryption (FHE) or secure multi-party computation (MPC). We first discuss the evaluation of non-confidential expressions. For brevity we only discuss the evaluation of expressions involving numbers, additions and multiplications. We focus on evaluating postfix expressions. When using encrypted values, the expression remains confidential. That is to say, despite computing the expression we do not learn anything about the expression except its length. The key to achieve this is the conditional push, ie. we execute a push operation but it only has an impact given that the element to be pushed is different from the special element $\varnothing$ that is not appended to the stack. Our algorithm \ref{alg:textexpr} requires linear run-time in terms of the number of elements in the expression (or, more precisely, in the bound we get for the length of the expression).

\begin{algorithm}[!htp]
	\caption{\textbf{Case Study: PostFix expressions }}\label{alg:textexpr}

	\begin{algorithmic}[1]
		\REQUIRE{LIFO Queue $A$ of postfix symbols of length at most $n$}
		\ENSURE{Result of evaluation } 
		\begin{small}			
			\STATE $st := LIFO(s)$ \COMMENT{Choose number of SAs $s$ such that array can hold at least $n$ elements}
			\FOR{$i:=0 \textbf{ to } n-1$}
			\STATE $symb := A.pop(1)$
			\STATE $toPush := symb$ if $symb$ is a number else $\varnothing$
			\STATE st.push(toPush)
			\STATE $isAdd := 1$ if $symb="+"$ else 0
			\STATE resAdd := st.pop(isAdd) + st.pop(isAdd)
			\STATE $isMul := 1$ if $symb="*"$ else 0
			\STATE resMul := st.pop(isMul) $\cdot$ st.pop(isMul)
			\STATE $toPush := isMul\cdot resMul + (1-isMul)\cdot\varnothing$
			\STATE $toPush := isAdd\cdot resAdd + (1-isAdd)\cdot toPush$
			\STATE st.push(toPush)
			\ENDFOR
			\STATE \textbf{return} st.pop(1)
		\end{small}
	\end{algorithmic}
	\small
\end{algorithm}

To evaluate confidential expressions, all array elements of the input $A$ must be encrypted as well as variables that depend on the array elements. Variables that indicate array lengths or the number of operations do not have to be encrypted.\footnote{In Algorithm \ref{alg:textexpr} the values of $n$ and $s$  do not have to be encrypted. In the LIFO queue and its sub-procedures, $n_{pu}$, $n_{po}$,$q$,$o$,$mi$ and $s$ remain unencrypted. All other variables are encrypted.}  To this end, one can use any of the known scheme for computing on encrypted data such as FHE or MPC, eg. \cite{gentry09} or \cite{sch15a}. These schemes provide basic operations such as addition and multiplication that allow to construct other operations such as comparisons, subtractions and more. However, certain operations like accessing an array element using an encrypted index might occur linear overhead in the length of the array. In our case, we manage to keep the asymptotic running time, since we only have to directly substitute additions, subtractions, multiplication and comparisons operations.

\subsection{Stock Span Analysis}
The Stock Span Problem is motivated by financial analysis of stocks. The span of a stock's price on day $d$ is the maximum number of consecutive days until day $d$, where the price of the stock has been at most its price on $d$. The well-known text book solution is given in Algorithm \ref{alg:textstock} taking linear time in the number of prices $n$.\footnote{We adjusted it from \url{http://www.geeksforgeeks.org/the-stock-span-problem/}} A straight forward solution gives a quadratic run-time algorithm due to the nested loops, ie. due to the worst case behavior of the inner loop. This renders the solution impractical for larger datasets. The total number of iterations (when being summed across all outer loop iterations) of the inner loop is only $n$. A single iteration of the inner loop could perform all $n$ iterations for some inputs. To ensure obliviousness we would have to execute (asymptotically) $n$ iterations of the inner loop for every execution of the outer loop. Furthermore, the code contains direct array access, eg. $price[i]$. In the obvious manner, this would also incur linear run-time overhead. However, it is possible to transform the nested loop by essentially changing the inner loop to an `if'-conditional first without changing the number of iterations of the outer loop. Then we make the loop oblivious using a conditional expression \texttt{if-then-else}. Essentially, in Algorithm \ref{alg:textstock} we replace the \texttt{while} and \texttt{do} keyword in line \ref{a:while} by an \texttt{if} and \texttt{then}. Lines \ref{a:then1} to \ref{a:then3} form the \texttt{else} part. We only show the final pseudocode after the translation of the `if' into oblivious code in Algorithm \ref{alg:obltextstock}. Since we must execute both branches of the $if$ to keep the condition confidential, the algorithm requires that we can execute the `pop' operation without impacting the data structure, ie. without actually performing a pop. This is supported by our data structure by using a special element in case the condition evaluates to true. The algorithm uses a peek operation, which returns the first element without removing it. It can be implemented using a combination of pop and push operation, eg. $x:=pop(1)$, $push(x)$.

\begin{algorithm}[!htp]
	\caption{\textbf{Case Study: Stock Span}}
	\begin{algorithmic}[1]
		\REQUIRE{ LIFO Queue $price$ of prices of length at most $n$}
		\ENSURE{ LIFO $S$ with spans (in reverse order)} \label{alg:textstock}
		\begin{small}			
			\STATE $st := LIFO(s)$ \COMMENT{Choose number of SAs $s$ such that array can hold at least $n$ elements}
			\STATE $st.push(0)$
			\STATE $S.push(1)$ \COMMENT{Span first element is 1}			
			\FOR{$i=0 \textbf{ to } n-1$}
			\STATE \textbf{while} {$st.peek()\neq \varnothing \wedge price[st[0]] \leq price[i]$} \textbf{do} st.pop() \label{a:while}
			\STATE $span := i+1$ if $st.peek()\neq \varnothing$ else $i-st[0]$ \label{a:then1}
			\STATE S.push(span)
			\STATE st.append(i) \label{a:then3}
			\ENDFOR
		\end{small}
	\end{algorithmic}
	\small
\end{algorithm}

\begin{algorithm}[!htp]
	\caption{\textbf{Case Study: Oblivious Stock Span }}\label{alg:obltextstock}

	\begin{algorithmic}[1]
				\REQUIRE{ LIFO Queue $price$ of prices of length at most $n$}
				\ENSURE{ LIFO $S$ with spans (in reverse order)} 
		\begin{small}			
			\STATE $st := LIFO(s)$ \COMMENT{Choose number of SAs $s$ such that array can hold at least $n$ elements}
			\STATE $pi:=price.pop(1)$			
			\STATE $st.push((0,pi))$
			\STATE $S.push(1)$ \COMMENT{Span first element is 1}	
			\STATE $i:=0$	
			\FOR{$k:=0 \textbf{ to } n-1$}
			\STATE (sti,stp) := st.pop(1)
			\STATE popNext := 1 if $sti = \varnothing$ or $sti\leq pi$ else 0
			\STATE $pi :=price.pop(popNext)\cdot popNext + (1-popNext)\cdot pi$
			\STATE $i:=i+(1-popNext)$
			\STATE $span := i+1$ if $st.peek()\neq \varnothing$ else $i-sti$
			\STATE $span := \varnothing$ if popNext else span
			\STATE $S.push(span)$
			\STATE $pushi := (i,pi)$ if (1-popNext) else $(\varnothing,\varnothing)$
			\STATE $st.push(pushi)$			
			\ENDFOR
		\end{small}
	\end{algorithmic}
	\small
\end{algorithm}


\section{Related Work}
In 2009 Gentry \cite{gentry09} introduced a fully homomorphic encryption(FHE) scheme based on lattices. Since then the field of computing on encrypted data (and circuits) has evolved rapidly, as summarized in \cite{moo14}. All approaches for FHE are based on either addition and multiplication or XOR and AND. Secure computations can also be carried out using multiple parties, such that no party learns a secret (if it does not collude with other parties). Secure multi-party computation was introduced three decades ago \cite{yao86,Gol87} and is still subject to extensive research, eg. \cite{sch15a}. Both SMC and FHE can compute expressions beyond addition and multiplication  such as comparisons. These operations could be used as black boxes to make our algoritms work on encrypted data. \cite{gentry09,van10} mentioned that circuit privacy is achievable by adding a (large) noise vector of the encrypted 0. The original work on SMC \cite{yao86} already allowed to hide circuits using garbled circuits. Our approach also allows to achieve circuit privacy in a novel manner by hiding whether a certain operation really impacted the computation. Our work is not limited to circuits.
``Oblivious RAM'' (ORAM)\ci{gol87Obl} disguises access patterns by a client from a server using randomization. The original solution\ci{gol87Obl} is based on a hierarchy of buffers such that each level of the hierarchy consists of several blocks. One block per level is read and always written on the first level. For each level, some blocks are real and some are dummy containing random data. The original algorithm has been improved, eg. in a recent tree-based scheme\ci{stef13} each data block is stored somewhere in the tree, ie. following a specific path from the root. After an access minor reordering involving the accessed elements takes place, potentially, resulting in some data being sent to the client. Some schemes, eg.\ci{stef13, wan14}, trade-off performance (or memory consumption) and the risk for memory access errors. Oblivious data structures for ORAM covering arrays and queues are discussed in\ci{wan14}. They make use of data access locality encountered for arrays and queues. A single access to a parent node in the position map returns pointers to multiple children. They hide whether the operation is a read or write to a memory cell. However, assuming one knows that a write must occur in a function, one knows that some memory cell is modified. We do not use traditional ORAM techniques. Furthermore, in our scenario, knowing that a certain operation is performed, ie. a pop or push, still gives no hint whether the data structure was modified or not.

Other work designed oblivious data structures particularly for SMC, eg. \cite{kel14,tof11}. The work \cite{kel14} uses ORAM structures and secret sharing among parties to achieve obliviousness. In contrast, \cite{tof11} presents a deterministic scheme for priority queues using the bucket heap concept for priority queues\ci{bro04} coming with O($\log^2 n$) overhead. Bucket heaps partition memory into buffers of size $2^{2i+1}$ and signal blocks of size $2^{2i}$\ci{bro04}. Buckets store actual elements, whereas buffers store overflowing elements. Once a buffer is full it is merged into a bucket. \cite{tof11} adjusted this setting to use blocks of equal size. Our queue shares the common idea of organizing data in blocks of increasing size that is also found in other work, eg.\ci{mit14}. We differ from prior work\ci{mit14,tof11,bro04} in several aspects, eg. we perform a more fine-grained partitioning using multiple blocks, eg. in the view of \cite{tof11} we introduce one more level of partitioning for buffer and data blocks. We have also come up with a deterministic oblivious sequence of restructuring operations to handle empty and full blocks rather than counting the number of elements in the queue, eg. \cite{tof11}. In contrast to our work, prior work also does not hide the impact of an operation (ie. they do not hide the number of elements in a bucket), which is essential for securing control structures. Our fast B2B-FIFO queue introduces novel ideas such as block sharing not found in prior work. 

The paper\ci{agg10} shows how to compute the $k$-th ranked element for SMC. The paper\ci{lin09} discusses median computation. Such operations might prove valuable also for sorting, eg. selecting the median element for quicksort. However, both protocols\ci{agg10,lin09} disclose the outcome of secure comparisons, which might require non-desirable client interaction and is not considered secure. The SMC protocol for sorting in \cite{zha11} runs in constant rounds but needs to know the product of the range of numbers $R$ and it has communication and computational complexity that is proportional to product of the range of numbers times the number of elements, ie. $O(n\cdot R)$ (an improved version has $O(n^2)$). To achieve constant rounds it relies on the evaluation of unbounded fan-in gates.

Sorting networks are naturally oblivious, since they use a fixed sequence of comparisons among elements in an array that is not related to the data stored in the array. They have been used for secure sorting in\ci{jon11,goo14}. The work\ci{goo14} is based on a network with $19600\cdot n\log n$ comparators. A comparator can be implemented by a comparison yielding a bit $b$ followed by an exchange of two elements $A,B$, ie. $A := b\cdot B + (1-b)\cdot A$ and $B := b\cdot A + (1-b)\cdot B$. Therefore a comparator needs 7 E-Ops in addition to the comparison, yielding $156800\cdot n\log n$ operations. Though this is asymptotically optimal, it is of little practical relevance due to the number of comparators needed. Additionally, the depth (of the sorting network) is of order $n\log n$, which makes it non-parallelizable. Our algorithms improve on it for all relevant scenarios (see Section \ref{sec:msort} for a detailed comparison). The oblivious randomized Shellshort algorithm \cite{goo10} is asymptotically optimal in terms of the number of comparisons using several techniques such as permutation of the array as well as shaker and brick passes of the array.

Oblivious algorithms for geometric problems are presented in \cite{epp10}. Algorithms for graphs incurring overhead up to linear factor (in the number of nodes) are given in \cite{bla13}. Other work \ci{roc15} based on ORAM designed data structures for maps. They allow for history independence, ie. different sequences of operations lead to indistinguishable (memory layouts of the physical) data structures. 

\section{Evaluation} \label{sec:eval}
We shed light on two aspects that are not immediate from the asymptotic analysis. First, on the one hand our oblivious data structures are more involved than using a naive oblivious implementation traversing the entire array for each operation, on the other hand we have shown that asymptotically it outperforms a naive implementation. The key question is, whether our oblivious queues outperform already for queues of small capacity or only for those with large capacity. Therefore, we compared our implementation against a simple `linear' oblivious queue that accesses all elements (that could be stored) for each operation. Thus, the run-time is linear in the capacity.
Second, how much slower is our array compared to a non-oblivious queue. We have shown that the asymptotic factors are of order O($\log n$) and O($\log^2 n$) depending on the queue type. Here, we give more precise insights.


We implemented the benchmarks in Python. The evaluation was run on a machine equipped with an Intel 2.5 GHz quad-core CPU with 8 GB RAM on Windows 7. For the non-oblivious queue we ran 1 Mio. operations. For the oblivous linear queue, ie. the naive oblivious queue traversing the entire array for each operation, we attempted to run 1 Mio operations, but stopped after 1 hour if the computation was still ongoing and estimated the time it would take to compute 1 Mio. operations. For the oblivious data structures we executed $\max(100000,2\cdot capacity)$ operations, since the maximal run-time is achieved if we execute a multiple of the capacity.  Each operation was chosen randomly among a push and pop operation. Due to the obliviousness it does not matter what parameters we use for the push and pop operation. 

\begin{figure*}[!htp]
	\begin{subfigure}{.45\textwidth}
		\centering
		\centerline{\includegraphics[width=\linewidth]{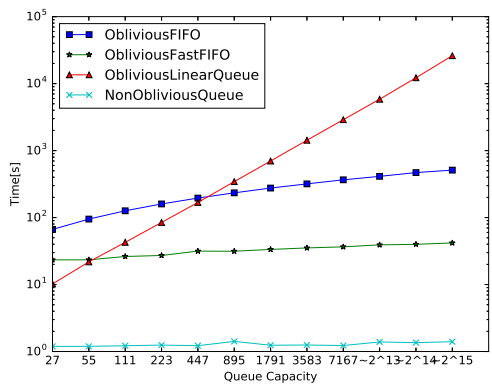}}
		\caption{FIFO Queues}
		\label{fig:fifo}
	\end{subfigure}%
	\begin{subfigure}{.45\textwidth}
		\centering
		\centerline{\includegraphics[width=\linewidth]{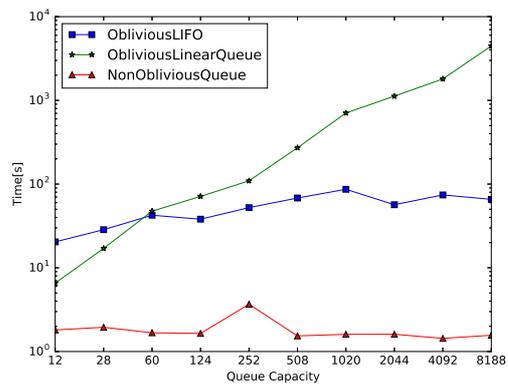}}
		\caption{LIFO Queues}
		\label{fig:lifo}
	\end{subfigure}
	
	\hfill
	
	\begin{subfigure}{.45\textwidth}
		\centering
		\centerline{\includegraphics[width=\linewidth]{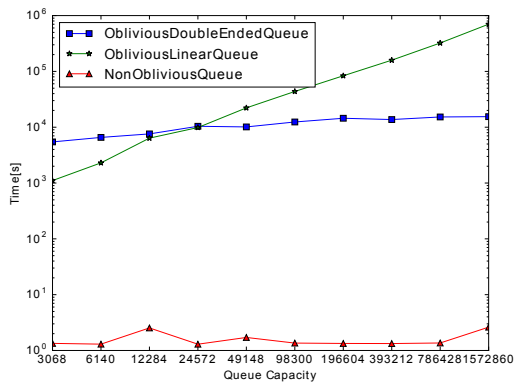}}
		\caption{Double-Ended Queues}
		\label{fig:array}
	\end{subfigure}
	\begin{subfigure}{.45\textwidth}
		\centering
		\centerline{\includegraphics[width=\linewidth]{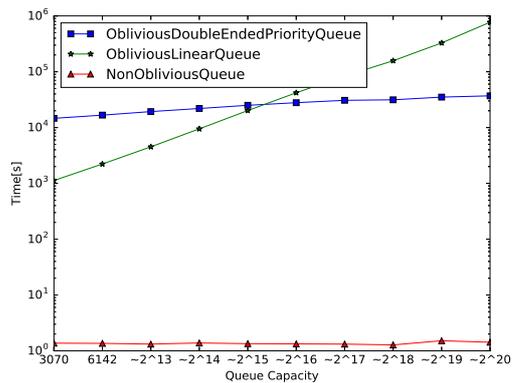}}
		\caption{Double-Ended Priority Queues}
		\label{fig:arrayPrio}
	\end{subfigure}
	\caption{Running Times results for 1 Mio. operations for our oblivious queues compared to linear oblivious and non-oblivious queues}
	\label{fig:fig}
\end{figure*}

The plots in Figures \ref{fig:fifo}, \ref{fig:lifo},\ref{fig:array} and \ref{fig:arrayPrio} show the run-times comparing all queue variants for increasing maximum queue sizes for FIFO and LIFO queues. Qualitatively all queues behave similarly as predicted by the asymptotic analysis. For small queue sizes (LIFO and FastFIFO up to about 60, FIFO up to about 500) a simple linear oblivious queue has an edge over our more complex queues. For double-ended queues performance is somewhat worse, but simple linear queues are also outperformed for moderate queue sizes. With growing queue sizes the exponential gap becomes clearly visible between the linear oblivious queue and our implementations. The LIFO and fast FastFIFO queue are more than 100x faster for queues of capacity about 10000. For FIFO queues we reach the boundary of 100x performance improvement for queues of capacity about 100000. 
Note, that it is not uncommon that a simple algorithm with bad asymptotic behavior outperforms more complex algorithms. For example, a naive bubble sort might also be faster than mergesort for small data sizes although the later is asymptotically much faster.\\

When it comes to overhead of our oblivious queue compared to the built-in Python queue (surrounded by a wrapper), which uses memory proportional to the actual array size, the asymptotic behavior is well-visible. Our LIFO and FastFIFO queue both have an asymptotic overhead of $\log n$ compared to non-oblivious queues that directly accesses queue elements. This results in close to parallel lines in Figures \ref{fig:lifo} and \ref{fig:lifo}. The overhead is roughly a factor 40 for queues of size 10000.  For FIFO queues the asymptotic overhead is larger, ie. $\log^2 n$. The overhead is a factor of 200 for arrays of the same size. In the light of overhead that typically comes with secure computation, eg. FHE or SMC that can reach more than 5-6 orders of magnitude \ci{nae11}, our overhead is very modest. We also want to emphasize that to the best of our knowledge no prior work has compared against a non-oblivious implementation. We believe this is a very important benchmark.

\section{Conclusions}
We have presented oblivious queues accompanied by theoretical and practical investigation having only  (poly)logarithmic overhead. Since queues are an essential part of everyday programming, we believe that they will play a major role for enabling computation using encrypted data, in particular with focus on expression hiding. Still, many more common data structures and operations have to be realized efficiently before any of the existing technologies such as FHE and MPC become practical for a large range of applications

\bibliographystyle{abbrv}
{\small
\bibliography{refs}}

\end{document}